\documentclass[12pt]{article}
\usepackage{epsf}
\title{Improved Calculations of Quark Distributions\\
in Hadrons: the case of pion.}
\vspace{2cm}
\author{B.L.Ioffe and A.G.Oganesian\\
\\
Institute of Theoretical and Experimental Physics\\
B.Cheremushkinskaya 25, 117218, Moscow, Russia}
\date{}
\begin{document}

\maketitle

\newcommand{\be}{\begin{equation}}
\newcommand{\ee}{\end{equation}}

\def\la{\mathrel{\mathpalette\fun <}}
\def\ga{\mathrel{\mathpalette\fun >}}
\def\fun#1#2{\lower3.6pt\vbox{\baselineskip0pt\lineskip.9pt
\ialign{$\mathsurround=0pt#1\hfil##\hfil$\crcr#2\crcr\sim\crcr}}}

\begin{abstract}
The earlier introduced method of calculation of quark distributions in
hadrons, based on QCD sum rules, is improved. The imaginary part of the
virtual photon forward scattering amplitude on some hadronic current is
considered in the case, when initial and final  virtualities of the current
$p^2_1$, and $p^2_2$  are different, $p^2_1\not= p^2_2$. The operator product
expansion (OPE) in $p^2_1$, $p^2_2$ is performed.  The sum rule for quark
distribution is obtained using double dispersion representation of the
amplitude on one side in terms of calculated in QCD OPE and on the other
side in terms of physical states contributions. Double Borel transformation
in $p^2_1$, $p^2_2$ is applied to the sum rule, killing background
non-diagonal transition terms, which deteriorated the accuracy in previous
calculations.  The case of valence quark distribution in pion is considered,
which was impossible to treat by the previous method. OPE up to dimension 6
operators is performed and leading order perturbative corrections are
accounted. Valence $u$-quark distribution in $\pi^+$  was found at
intermediate $x$, $0.15 < x < 0.7$  and normalization point $Q^2=2~GeV^2$.
These results may be used as input for evolution equations.
\end{abstract}

\section{Introduction}

The QCD sum rule approach, invented by Shifman, Vainstein and Zakharov in
1979 [1]  is now  well known as a powerful method, which make possible to
calculate in QCD in non-model way and with a good accuracy various hadron
characteristic like masses, decay widths, formfactors etc. The method is
based on the operator product expansion (OPE), extended to the
nonperturbative region.
These results were obtained from consideration of 2 and 3-point correlators
(for review see \cite{2}). A bit later the structure functions -- quark
distribution in photon and hadrons were investigated in the QCD sum rule
framework. The second moment of photon structure function was considered in
\cite{3}, and for pion and nucleon -- in \cite{4,5}, but unfortunately it
was difficult to extend this approach for calculating higher moments. The
general method how to calculate hadron structure functions in the region of
intermediate $x$  was suggested in \cite{6} and developed in \cite{7}. The
method is based on the consideration of 4-point correlator, corresponding to
forward scattering of two currents, one of which has quantum number of
hadron of interest, and the other is electromagnetic (or weak).
In the first order of OPE, in the case,
when the hadron is a meson, this corresponds to box diagrams like  shown in
Fig.1, where $p$  is momentum of hadron current and $q$  is momentum of
photon. The problem of such diagrams is that even if $p^2$, $q^2$ are large
and negative, in the case of forward scattering
the singularity in $t$-channel for massless quarks is at $t=0$, i.e. large
distances in $t$-channel are of importance. However, as was shown in
\cite{6,7}  the situation changes drastically when the imaginary part of the
scattering amplitude -- the object of interest in case of structure
functions -- is considered. The imaginary part in $s$-channel $(s=(q+p)^2)$
of the forward scattering amplitude is dominated by small distances
contributions at large (negative)  $p^2$ and intermediate $x$. (Here the
standard notation is used: $x$ is the Bjorken scaling variable,
$x=-q^2/2\nu$, $\nu=p q$).  The proof of this statement, given in \cite{7},
is based on the fact, that for the imaginary part of the forward
amplitude the position of closest to zero singularity in momentum transfer
is determined by the boundary of the Mandelstam
spectral function and given by the equation

%1
\be
t = -4\frac{x}{1-x} p^2
\label{1}
\ee
(it is assumed, that $\mid q^2 \mid \gg \mid p^2\mid$).
Therefore, even at $t=0$, but not small $x$  and large $p^2$  the
virtualities of intermediate states in the $t$-channel are large enough for
OPE be available. The further procedure is common for QCD sum rule (with some
special nuances we will discuss later), i.e. dispersion representation on
$p^2$ is saturated by physical states and the contribution of the lowest
particle state is extracted using Borel transformation. In \cite{7}  the
structure function of nucleon was calculated. Somewhat later, structure
function of photon has also been calculated \cite{8}. But one should note,
that sum rule for $d$-quark distribution in proton obtained in \cite{7}, is
applicable within rather narrow range of $x$ $(0.2 < x < 0.45)$  and
agreement with experiment is not good enough. Moreover, it was found to be
impossible to calculate structure functions of $\pi$- and $\rho$-mesons in
this way (that's why the authors of \cite{8} was forced to use special
trick, based on VDM, to calculate $\rho$-meson structure function). The
reason for this is that the sum rules, in form used in \cite{7}, have a
serious drawback.

To understand, what kind of problem it is, let us shortly review the main
points of the method. Let us consider 4-point correlator with two
electromagnetic currents and two currents with quantum number of some hadron
(for clarity the axial current, corresponding to charged pions, will be
considered but conclusion is independent of the choice of current):

%2
$$\Pi_{\mu\nu\lambda\rho}(p_1,p_2;q_1,q_2) = -\int
e^{ip_1x+iq_1y-ip_2z}d^4xd^4yd^4z\langle 0\mid T \left \{j_{5\lambda}(x)
j^{em}_{\mu}(y)j^{em}_{\nu}(0)j_{5\rho}(z)\right \}\mid 0 \rangle $$
\be
j_{5\lambda}=\bar{u}\gamma_5\gamma_{\lambda}d
\label{2}
\ee
By considering of forward scattering amplitude in accord with \cite{7}, put
$p_1=p_2$  at the very beginning.
Among various tensor structures of $\Pi_{\mu\nu\lambda\rho}$ it is
convenient to consider the structure
$(p_{\mu}p_{\nu}p_{\lambda}p_{\rho}/\nu)\cdot\tilde{\Pi}(p^2,q^2,x)$, and
the imaginary part $Im~\tilde{\Pi}(p^2,q^2,x)$ in $s$-channel  is related 
to pion structure function $F_{2\pi}(x)$\footnote{As was mentioned 
in \cite{7}, the results are more reliable, if invariant amplitude at 
kinematical structure with maximal dimension is used.}. Let us write 
dispersion relation representation of $Im~\tilde{\Pi}(p^2,q^2,x)$ in the 
$p^2$ variable. As was shown in \cite{7} (see also \cite{9},\cite{10}) the 
correct form of dispersion representation is double dispersion relation

%3
\be
Im~\tilde{\Pi} = a(x) + \int\limits^{\infty}_0
\frac{\varphi(x,u)du}{(u-p^2)} + \int\limits^{\infty}_0
\int\limits^{\infty}_0
\frac{dudu^{\prime}\rho(u,u^{\prime},x)}{(u-p^2)(u^{\prime}-p^2)}
\label{3}
\ee
(We consider lowest twist contributions, the terms of order
$p^2/q^2$ are neglected.) In order to derive (\ref{3}) it is
convenient to consider first the case, when $p^2_1\not= p^2_2$ and go to the
limit $p^2_1\to p^2_2=p^2$. Then the form (\ref{3}) is evident. The last
term in the right  hand side (rhs) of (\ref{3})  represents the propertly
double dispersion contribution, the second may be considered as subtraction
term in variables $p^2_1$  or $p^2_2$  and the first term arises as
subtraction from the second. The interesting for us contribution arises from
the pion poles in both variables -- $u$  and $u^{\prime}$  in the last term
in (\ref{3}). This term corresponds to the diagram of Fig.2, where the axial
current creates the pion, then the process of deep inelastic scattering of
virtual photon on pion proceeds and finally pion  is absorbed by axial
current. Evidently this term is proportional to the pion structure function.
All others in (\ref{3})  may be considered as background.
Accept a model of hadronic spectrum, in which
$\rho,\varphi$  can be represented by contribution of resonance
($\pi$-meson) and continuum ($s_0$ is continuum threshold)

%4
$$
\rho(u,u^{\prime},x) = f(x)\delta(u-m^2_{\pi})\delta(u^{\prime}-m^2_{\pi})
+\rho^0(x)\theta(u-s_0)\theta(u^{\prime}-s_0)$$
\be
\varphi(x,u) =  \varphi_1(x)\delta(u-m^2_{\pi}) +\varphi_2(x) \theta(u-s_0)
\label{4}
\ee
where $f(x)$  is proportional to resonance ($\pi$-meson)  structure function
of interest,

%5
\be
f(x) \sim 2\pi F_2(x)
\label{5}
\ee
and $\varphi_{1,2}$  are some unknown functions, corresponding to
non-diagonal transitions.

The substitution of (\ref{4}), into (\ref{3}) gives

%6
\be
I m~\tilde{\Pi} = \frac{f(x)}{(p^2-m^2_{\pi})^2} + a(x) +
\int\limits^{\infty}_{s_0} \int\limits^{\infty}_{s_0}
\frac{\rho^0(x,u,u^{\prime})dudu^{\prime}}{(u-p^2)(u^{\prime}-p^2)}  +
\int\limits^{\infty}_{s_0}\frac{\varphi_2(x,u)}{(u-p^2)}du +
\frac{\varphi_1(x,m^2_{\pi})}{(p^2-m^2_{\pi})}
\label{6}
\ee
The last term in (\ref{6}) corresponds to Fig.3, where axial current creates
a pion, deep inelastic scattering proceeds, but the final state is not a pion
like in Fig.2, but some excited state with pion quantum numbers, which is
absorbed by axial current. In order to separate the term proportional to the
pion structure function -- the first term in the rhs of (\ref{6}), the Borel
transformation in $p^2$ is applied to (\ref{6}), which suppresses continuum
contributions to (\ref{6}). (The Borel parameter $M^2$ is chosen such that
$e^{-s_0/M^2} \ll 1$). After Borel transformation we get:

%7
$$
{\cal{B}}_{M^2}~Im~ \tilde{\Pi} (p^2, x) = f(x)
\frac{1}{M^2}~e^{-m^2_{\pi}/M^2} - \varphi_1(x)~e^{-m^2_{\pi}/M^2} +
$$
\be
+ \int\limits^{\infty}_{s_0}~ du~ \int\limits^{\infty}_{s_0}~ du^{\prime}
\rho (x, u, u^{\prime}) e^{-(u + u^{\prime})/M^2} +
\int\limits^{\infty}_{s_0} \varphi_2 (x, u) e^{-\frac{u}{M^2}}
\label{7}
\ee
For the last two terms in rhs of (\ref{7}), we can assume that
$\rho(x,u,u^{\prime})$  and $\varphi_2(x,u)$ are given by contribution of
bare loop -- Fig.1. Because of Borel suppression $\sim exp(-s_0/M^2)\ll 1$
these terms are small and such an approximation does not introduce an 
essenthial
error in the final result. However, the second term in rhs of (\ref{7})  is
not exponentially suppressed in comparison with the first. The only way to
kill it is to differentiate both sides of (\ref{7}) (multiplied by
$exp(m^2_{\pi}/M^2)$) over $1/M^2$. Just this procedure was used in
\cite{6,7}  to determine nucleon structure functions. But, as is well known,
the differention of approximate relation may seriously deteriorate the
accuracy of the results. In QCD sum rules
such procedure increases contribution of nonperturbative corrections and
continuum contributions, sum rules become much worse or even fails (as for
$\rho$-meson). For $\pi$-meson situation is even worse, because direct
calculations show, that bare loop contribution corresponds only to
non-diagonal transitions.

In this work we suggest the modified method of
calculation of the hadron structure function, which is free  from this
problems and is completely based on QCD sum rules. We will illustrate it on
an example on the $\pi$-meson structure function calculation, which usually
is much "dangerous" case.

\section{The idea of the method.}

The idea of the method is to consider at the begining non-equal $p^2_1\not=
p^2_2$  in (\ref{2})  and perform all calculations for this case. Instead
(\ref{3}) dispersion representation takes the form

%8
\be
Im~\tilde{\Pi}(p_1^2,p^2_2,x) = a(x) + \int\limits^{\infty}_0
\frac{\varphi(x,u)}{u-p^2_1} du +
\int\limits^{\infty}_0\frac{\varphi(x,u)}{u-p^2_2} du+
\int\limits^{\infty}_0
du_1\int\limits^{\infty}_0
du_2\frac{\rho(x,u_1,u_2)}{(u_1-p^2_1)(u_2-p^2_2)}
\label{8}
\ee
Apply to (\ref{8})  double Borel transformation in $p^2_1,p^2_2$. This
transformation kills three first terms in rhs of (\ref{7})  and we have

%9
\be
{\cal{B}}_{M^2_1} {\cal{B}}_{M^2_2} Im~\tilde{\Pi}(p^2_1,p^2_2,x) =
\int\limits^{\infty}_0
du_1\int\limits^{\infty}_0
du_2\rho(x,u_1,u_2)exp\Biggl [-\frac{u_1}{M^2_1} -
\frac{u_2}{M^2_2} \Biggr ]
\label{9}
\ee
One can divide the integration region over $u_{1,2}$ into 4 areas (Fig.4):

I ~~$u_1 < s_0; u_2 < s_o$;

II ~$u_1 < s_0$; $u_2 > s_0$

III $u_2< s_0$; $u_1 > s_0$

IV $u_{1,2} > s_0$

\noindent
Using the standard  QCD sum rule model of hadronic spectrum and the
hypothesis of quark-hadronic duality, i.e. the model with one lower
resonance plus continuum, one can easily notice\footnote{We restrict
ourselves by simplest model, because higher resonance contribution in any
case will be suppressed after double borelization.},
that area (I) corresponds to resonance region. Spectral density can be
written in this area as

%10
\be
\rho(u_1,u_2,x) = f^2_{\pi} \cdot 2\pi
F_2(x)\delta(u_1-m^2_{\pi})\delta(u_2-m^2_{\pi})
\label{10}
\ee
where $f_{\pi}$ is defined as

$$\langle 0\mid j_{\lambda 5} \mid \pi \rangle = f_{\pi}
p_{\lambda}~~~~f_{\pi}=131~MeV$$
In area (IV), where both variables $u_{1,2}$ are far from resonance region,
the non-perturbative effects may be neglected, and, as usual in sum rules,
spectral function of hadron state is described by the bare loop spectral
function $\rho^0$  in the same region

%11
\be
\rho(u_1,u_2,x) = \rho^0(u_1,u_2,x)
\label{11}
\ee
In the areas (II),(III) one of variables is far from resonance region, but
other is in the resonance region, and spectral function in this region is
some unknown function $\rho=\psi(u_1,u_2,x)$, which corresponds to
transitions like $\pi \to$  continuum, as shown in Fig.3. After double Borel
transformation the total answer for physical part can be written as
($M^2_1,M^2_2$ are Borel masses square)

%12
$$\hat{B}_1\hat{B}_2[Im~\Pi] = 2\pi F_2(x)\cdot
f^2_{\pi}e^{-m^2_{\pi}( \frac{1}{M^2_1}+\frac{1}{M^2_2})} +
\int\limits^{s_0}_0 du_1 \int\limits^{\infty}_{s_0} du_2
\psi(u_1,u_2,x)e^{-(\frac{u_1}{M^2_1}+\frac{u_2}{M^2_2})}$$
\be
+ \int\limits^{\infty}_{s_0}du_1 \int\limits^{s_0}_0 du_2
\psi(u_1,u_2,x)e^{-(\frac{u_1}{M^2_1}+\frac{u_2}{M^2_2})}
+ \int\limits^{\infty}_{s_0}\int\limits^{\infty}_{s_0} du_1du_2
\rho^0(u_1,u_2,x) e^{-(\frac{u_1}{M^2_1}+\frac{u_2}{M^2_2})} \label{12}
\ee
In what follows we put for simplicity $M^2_1= M^2_2\equiv 2M^2$.  The one of
advantages of this method is that after double Borel transformation unknown
contribution of (II), (III) areas (second and third term in (\ref{12})) are
exponentially suppressed. Using duality arguments  we estimate the
contribution of all non-resonance region (i.e areas II,III,IV)  as
contribution of bare loop in the same region and demand their value to be
small (less than 30\%). So, equating physical and QCD representation of
$\tilde{\Pi}$, and taking in account cancellation of appropriate parts in
left and right sides one can write the following sum rules (we omit all
terms, which are suppressed after Borel transformation)

%13
$$ Im~\Pi^0_{QCD} + \mbox{Power correction} = 2\pi
F_2(x)f^2_{\pi}$$
\be
Im~\Pi^0_{QCD}  = \int\limits^{s_0}_0 \int\limits^{s_0}_0 \rho^0
(u_1,u_2,x) e^{-\frac{u_1+u_2}{2 M^2}};
\label{13}
\ee
(The pion mass is neglected.)
It can be shown (see Appendix), that for box
diagram $\psi(u_1,u_2,x)\sim\delta(u_1-u_2)$ and, as a consequence, the
second and third terms in (\ref{12}) are zero in our model of hadronic
spectrum.

\section{Calculation of box diagram.}

The diagrams, corresponding to unit operator contribution, are shown in
Fig.1a,b. Note, that crossing diagram, shown in Fig.1c does not contribute,
their contribution found to be 0 in leading twist.
(This is a sequence of kinematics, so such crossing diagrams also are zero
for higher dimension corrections in the leading twist.)

It is enough for us to calculate the distribution of valence $u$-quarks in
pion, since $\bar{d}(x)=u(x)$. For this reason restrict ourselves by
calculation of $Im~\tilde{\Pi}$  for the diagram Fig.1a.

Consider first the case $p_1=p_2$  and demonstrate, as was announced in
Sec.2, that in this case the contribution of box diagram attributes only to
non-diagonal transitions, like in Fig.3 and refers to background terms in
(\ref{7}). Diagram Fig.1a contribution is equal

%14
$$Im \Pi_{\mu\nu\lambda\sigma} = -\frac{3}{(2\pi)^2} \frac{1}{2}\int
\frac{d^4 k}{k^4}\delta [~(k+q)^2]\delta [~(p-k)^2] \times$$
\be
\times Tr[~\gamma_{\lambda}\hat{k}\gamma_{\mu}(\hat{k}+\hat{q})\gamma_{\nu}
\hat{k} \gamma_{\sigma}(\hat{k}-\hat{p})~]
\label{14}
\ee
Calculate the trace and omit the terms, which cannot contribute to the
interesting for us structure $\sim p_{\mu}p_{\nu}p_{\lambda}p_{\sigma}/\nu$.
We get

%15
\be
Im~ \Pi_{\mu\nu\lambda\sigma} = -\frac{12}{\pi^2} \int \frac{d^4
k}{k^4}k_{\mu}k_{\nu}k_{\lambda}(k_{\sigma}-p_{\sigma})\delta[(k+q)^2]
\delta[(p-k)^2]
\label{15}
\ee
Calculation of the integral leads to:

%16
\be
Im~ \Pi_{\mu\nu\lambda\sigma} =
-\frac{3}{\pi}p_{\mu}p_{\nu}p_{\lambda}p_{\sigma}\frac{1}{\nu p^2}x^2 (1-x)
\label{16}
\ee
(only the terms $\sim p_{\mu}p_{\nu}p_{\lambda}p_{\sigma}$ are kept)  and

%17
\be
Im~ \tilde{\Pi}(p^2,x) = -\frac{3}{\pi}\frac{1}{p^2}x^2(1-x)
\label{17}
\ee
Substitute (\ref{17}) into (\ref{6})  and perform Borel transformation. We
get:

%18
\be
\frac{3}{\pi}x^2(1-x)(1-e^{-s_0/M^2}) = 2\pi f^2_{\pi}x
u_{\pi}(x)\frac{1}{M^2} + \varphi_1(x),
\label{18}
\ee
where $u_{\pi}(x)$  is the distribution of valence $u$ quarks in pion (pion
mass is neglected). Looking at $M^2$ dependence  in (\ref{18})  it becomes
evident, that in this appropach the attempt to separate the pion
contribution from the background by studying $M^2$  dependence (e.g.
differentiation over $1/M^2$) is useless -- up to small correction $\sim
e^{-s_0/M^2}$ the box diagram contributes to the background only.

Consider now the more promisable approach, $p^2_1\not= p^2_2$. Since
nonequality of $p^2_1$, $p^2_2$  is important for us only for Borel
transformation, i.e. in the denominators of dispersion representation
(\ref{8}), in the calculation of numerator, resulting in kinematical
structure $p_{\mu}p_{\nu}p_{\lambda}p_{\alpha}$  we can put $p_1=p_2=p$.
Therefore, in order to understand the essential features of corresponding
integrals in case of non-equal $p^2_1,p^2_2$, it is sufficient to study
insread of (\ref{14}) a more simple integral

%19
\be
Im~T(p^2_1,p^2_2,q^2,\nu) = \int d^4
k\frac{1}{k^2}\frac{1}{(k+p_2-p_1)^2}\delta [~(k+q)^2]\delta [~(p_1-k)^2]
\label{19}
\ee
The direct calculation of the integral in rhs of (\ref{19}) (see Appendix)
shows, that it may be represented in the form

%20
\be
Im~T(p^2_1,p^2_2,q^2, \nu) = \frac{\pi}{4\nu x} \int\limits^{2\nu/x}_0
\frac{1}{u-p^2_1} \frac{1}{u-p^2_2} du
\label{20}
\ee
(Higher order terms in $p^2_1/q^2$, $p^2_2/q^2$ are neglected.) At
$p^2_1=p^2$  it gives

%21
\be
Im~T(p^2,q^2,\nu) = \frac{\pi}{4\nu x p^2}
\label{21}
\ee
as it should be. (\ref{20})  may be rewritten in the form of double
dispersion representation (\ref{8}) with $a(x)=\varphi(x)=0$  and
$\rho(u,u^{\prime},x)$ proportional to $\delta(u-u^{\prime})$

%22
\be
\nu Im~T(p^2_1,p^2_2,x) = -\frac{\pi}{4x}\int\limits_0^{\infty}
\frac{\delta(u-u^{\prime})}{(u-p^2_1)(u^{\prime}-p^2_2)} du~ du^{\prime}
\label{22}
\ee
(Higher twist terms are omitted). From this consideration it becomes clear,
that in order to go from the case of $p^2_1=p^2_2=p^2$  in the calculation
of the box diagram Fig.1a (\ref{14}) to $p^2_1\not=p^2_2$, it is enough to
substitute in the final result the factor $1/p^2$ by\footnote{It must be
mentioned, that such substitution is valid only for box diagram, it does not
take place for more complicated diagrams, considered in next Section.}

%23
\be
\frac{1}{p^2} \to -\int\limits^{\infty}_0 du \int\limits^{\infty}_0
du^{\prime} \frac{\delta(u-u^{\prime})}{(u-p^2_1)(u^{\prime}-p^2_2)}
\label{23}
\ee
Therefore instead of (\ref{17})  we get

%24
\be
\tilde{\Pi}(p^2_1,p^2_2,x) = \frac{3}{\pi} x^2(1-x)
\int\limits^{\infty}_0du  \int\limits^{\infty}_0  du^{\prime}
\frac{\delta(u-u^{\prime})}{(u-p^2_1)(u^{\prime}-p^2_2)}
\label{24}
\ee
Perform double Borel transformation in $p^2_1,p^2_2$. It kills nondesirable
depending on one variable subtraction terms in (\ref{8})  and we have the
sum rule for valence $u$-quark distribution in pion

%25
\be
u_{\pi}(x) = \frac{3}{2\pi^2}\frac{M^2}{f^2_{\pi}}x (1-x)(1-e^{-s_0/M^2}),
\label{25}
\ee
where it was put $M^2_1=M^2_2=2M^2.$
(As is known [11] the charactiristic values of Borel parameters
$M^2_1,~M^2_2$ in double Borel transformation are about twice of Borel
parameters in ordinary Borel transformation, used in mass calculations.)

Before going to more accurate consideration with account of higher dimension
operators and leading order (LO)  perturbative corrections, let discuss the
unit operator contribution in order to estimate, if it is reasonable. The
calculation of the pion decay constant $f_{\pi}$, performed in \cite{1}, in
the same approximation results in

%26
\be
f^2_{\pi} = \frac{1}{4\pi^2} M^2(1-e^{-s_0/M^2})
\label{26}
\ee
Substitution (\ref{26}) into (\ref{25}) gives

%27
\be
u_{\pi}(x) = 6 x(1-x)
\label{27}
\ee
One can note, that

%28
\be
\int\limits^1_0 u_{\pi}(x)dx = 1
\label{28}
\ee
in agreement with the fact, that in the quark-parton model there is one
valence quark in pion. Also, it can be is easily verified, that

%29
\be
\int\limits^1_0 x u_{\pi}(x)dx = 1/2
\label{29}
\ee
what corresponds to naive quark model, where no sea quarks exist. So one can
say, that formally unit operator contribution corresponds to naive parton
model.

Of course, eq.(\ref{28})  has only formal sense, because, as was discussed
in Introduction, our approach is correct only in some intermediate region of
$x$. The boundaries of $x$, where this approach is correct, will be found,
if one takes into account nonperturbative power corrections. In the next
section we will discuss them. At the end of this section let us discuss
perturbative corrections. We take into account only LO terms, proportional
to $ln(Q^2/M^2)$, and choose $Q^2=Q^2_0 \simeq 2 GeV^2$  -- for the point we
calculate our sum rules.
Finally, the result for bare loop has the form (the second term in square
brackets corresponds to perturbative correction, taken into account).

%30
\be
u_{\pi}(x) = \frac{3 M^2 x}{2 \pi^2 f^2_{\pi}} (1 - x) \Biggl [1
+ \frac{\alpha_s(\mu^2) ln (Q^2/\mu^2)}{3 \pi} (1/x + 4 ln(1 - x) -
\frac{2(1 - 2x) ln x}{1 - x}) \Biggr ] \cdot
(1 - e^{-s_0/M^2})
\label{30}
\ee
In the calculation we choose the normalization point $M^2 = \mu^2$.
The fact that we take into accounrt $\alpha_s$-correction at the point
$Q^2 = 2~ GeV^2$ means that our final results for the structure function
(we write it in the next section) can be used as an input for
evolution, starting from this value of $Q^2_0$.

\section{Calculations of higher order terms in OPE.}

In this section we discuss the power correction contribution to sum
rules. The power correction with lower dimension is proportional to gluon
condensate $\langle G^q_{\mu \nu} G^q_{\mu \nu} \rangle$ with $d=4$. As was
discussed above, only s-channel diagrams (Fig.1a) exists in the case of
double borelization. $\langle G^q_{\mu \nu} G^q_{\mu \nu} \rangle$
correction was calculated in a standard way in the Fock-Schwinger gauge
$x_{\mu}A_{\mu} = 0$ \cite{12}.

The quark propagator $iS(x,y) = \langle \psi(x) \bar{\psi}(y) \rangle$
in the external field $A_{\mu}$ has the well known form \cite{13}
(our sign of $g$ is opposite to that of \cite{13}):

%31
$$
iS(x,y) = iS^0(x-y) - g \int d^4 z iS^0(x-z) \cdot i\hat{A}(z)
iS^0(z-y) +
$$
\be
g^2 \int\int d^4z d^4z^{\prime}~iS^0(x-z) i\hat{A}(z) iS^0(z-z^{\prime})
\cdot i\hat{A}(z^{\prime}) \cdot iS^0(z^{\prime}-z) + ...
\label{31}
\ee
Here $S^0$ is free quark propagator; $\hat{A} =
(1/2) \lambda^a \gamma_{\mu} A^a_{\mu}$ and

%32
\be
A^a_{\mu}(x) = (\frac{1}{2}) x_{\rho} G^a_{\rho \mu} + (\frac{1}{3})
x_{\alpha} x_{\rho} [D_{\alpha} G^a_{\rho \mu}(0)] + (\frac{1}{8})
x_{\alpha}x_{\rho}x_{\beta}[D_{\alpha} D_{\beta} G^a_{\rho \mu} ]
\label{32}
\ee
When calculating one should take into account quark propagator
expansion up to the third term and only the first term in the expansion
of the external field $A_{\mu}$ (Fig.5).

These diagrams have been calculated using the program of analytical
calculation REDUCE. Surprisingly, in the case of the double
borelization the sum of all diagrams Fig.5 was found to be 0. So, the
gluon condensate contribution to the sum rule is absent.

Before we discuss the $d=6$ contribution, let us make the following
remark. Due to the fact that we are interested only in the intermediate
values of $x$, we should take into account only loop diagrams. Really,
one can easily see, that the diagrams with no loops (like those in
Fig.6) are proportional to $\delta(1-x)$ and is out of the region of the
method applicability. There are a large number of loop diagrams,
corresponding to $d=6$ corrections. First of all, there are diagrams
which correspond to interaction only with gluon vacuum field, i.e. only
with external soft gluon lines (see Fig.7). Such diagrams may
appear, if we take:

a) all possible combinations, which appear when expansion of  quark
propagator (\ref{31}) is taken into account up to the fourth term and
in expansion of the external gluon field (\ref{32})
 only the first term is kept. For example, it is the fourth term of the
expansion for one quark propagator and the first term (free propagator)
for other three (Fig.7a), the second term of the expansion for three
quark propagator and one propagator is free (Fig.7b), the third term 
of the expansion for one quark propagator and the second term for 
other (Fig.7c) etc.

b) all possible combination, when the second and the third terms of
expansion of gluon field (\ref{32})  is taken into account, like those,
shown in Fig.8. 

The diagrams of Fig.7 are, obviously, proportional to $\langle g^3
f^{abc} G^a_{\mu \nu} G^b_{\alpha \beta} G^c_{\rho \sigma} \rangle$ and
when calculating it is convenient to use representation of this tensor
structure given in \cite{14}

%33
$$
\langle 0 \mid g^3 f^{abc} G^a_{\mu \nu} G^b_{\alpha \beta} G^c_{\rho
\sigma} \mid 0 \rangle = (1/24) \langle 0 \mid f^{abc} G^a_{\gamma \delta
}G^b_{\delta \epsilon} G^c_{\epsilon \gamma} \mid 0 \rangle \cdot
(g_{\mu \sigma}g_{\alpha \nu} g_{\beta \rho} +
$$

$$
+ g_{\mu \beta} g_{\alpha \rho} g_{\sigma \nu} + g_{\alpha \sigma} g_{\mu
\rho} g_{\nu \beta} + g_{\rho \nu} g_{\mu \alpha} g_{\beta \sigma} -
$$

\be
- g_{\mu \beta} g_{\alpha \sigma} g_{\rho \nu} - g_{\mu \sigma} g_{\nu
\beta}g_{\alpha\rho} - g_{\alpha \nu} g_{\mu \rho} g_{\beta \sigma} -
g_{\beta \rho} g_{\mu \alpha} g_{\nu \sigma} )
\label{33}
\ee
The diagrams
of Fig.8 are proportional to $\langle 0 \mid D_{\rho} G^a_{\mu \nu} D_{\tau}
G^a_{\alpha \beta} \mid 0 \rangle$ and \\
$\langle 0 \mid  G^a_{\mu
\nu} D_{\rho} D_{\tau }G^a_{\alpha \beta} \mid 0 \rangle$ .  Using the
equation of motion it was found in \cite{14}

%34
$$
-\langle 0 \mid D_{\rho} G^a_{\mu \nu} D_{\sigma} G^a_{\alpha \beta}
\mid 0 \rangle = \langle 0 \mid D_{\rho}
G^a_{\mu \nu} D_{\rho} D_{\sigma }G^a_{\alpha \beta} \mid 0 \rangle =$$

$$
= 2 O^- \Biggl [g_{\rho \sigma} (g_{\mu \beta}g_{\alpha \nu} - g_{\mu
\alpha} g_{\nu \beta}) + \frac{1}{2} (g_{\mu \beta} g_{\alpha \sigma}
g_{\rho \nu} + g_{\alpha \nu} g_{\mu \rho} g_{\beta \sigma} - g_{\alpha
\sigma} g_{\mu \rho} g_{\nu \beta} - g_{\rho \nu} g_{\mu \alpha} g_{\beta
\sigma}) \Biggr ] +
$$
$$
O^+(g_{\mu \sigma} g_{\alpha \nu} g_{\beta \rho} + g_{\mu \beta}
g_{\alpha \rho} g_{\sigma \nu} - g_{\mu \sigma} g_{\alpha \rho} g_{\nu
\beta} - g_{\rho \beta} g_{\mu \alpha} g_{\nu \sigma});
$$
\be
O^{\pm} = \frac{1}{72} \langle 0 \mid g^2 j^a_{\mu} j^a_{\mu} \mid 0
\rangle \pm \frac{1}{48} \langle 0 \mid g f^{abc} G^a_{\mu \nu}
G^b_{\nu \lambda} G^c_{\lambda \mu} \mid 0 \rangle
\label{34}
\ee
where $j^a_{\mu} = \sum \limits_{i} \bar{\psi}_i \gamma_{\mu}
\frac{\lambda^a}{2}~\psi_i$.

From (\ref{33}), (\ref{34}) one may note, that these tensor structure
are proportional to two different vacuum averages:

$$
\langle 0\mid g^2 j^2_{\mu}\mid 0\rangle ~~ \mbox{and} ~~
\langle 0 \mid
g^3 G^a_{\mu \nu} G^b_{\nu\rho} G^c_{\rho \mu} f^{abc} \mid 0 \rangle
$$
First of them $\langle 0\mid g^2 j^2_{\mu}\mid 0\rangle$ by use of the
factorization hypothesis easily reduces to $\langle g \bar{\psi}\psi
\rangle^2$, which is well known.

\be
\langle 0\mid g^2 j^2_{\mu}\mid 0\rangle = -(4/3) [\langle 0\mid g
\bar{\psi}\psi\mid 0\rangle ]^2
\label{35}
\ee
But $\langle 0 \mid
g^3 G^a_{\mu \nu} G^b_{\nu\rho} G^c_{\rho \mu} f^{abc} \mid 0 \rangle
$ is not well known, there
are only some estimations based on the instanton model \cite{15},
\cite{16}. Fortunately, in the sum of all diagrams of this two types 
all terms proportional to this dimension 6 gluonic condensate are
exactly cancelled, and the sum of diagrams of Fig.7 and Fig.8
is proportional only to $\langle g \bar{\psi} \psi \rangle^2$.

We consider now an another type of diagrams which also give
contribution to $d=6$ power corrections. Such diagrams appear when the
external quark field is taken into account, i.e., one should take into
consideration the expansion of quark field:

\be
\psi(x) = \psi(0) + x_{\alpha_1} [\nabla_{\alpha_1} \psi(0)] +
\frac{1}{2} x_{\alpha_1}x_{\alpha_2}[\nabla_{\alpha_1}
\bar\nabla_{\alpha_2} \psi(0)] + ...
\label{36}
\ee
where $\nabla$ is covariant derivative.

In this case there appear diagrams like those in Figs.9-11, where quark
(and antiquark) line is expanded and the first and the second terms of
the expansion (\ref{36}) are taken. The expansion of the external gluon
field (\ref{32}) is also accounted up to the second term. For the
diagrams of Fig.10 gluon propagator in the external field is also
accounted (we discuss it a bit later).

All these diagrams can be divided into two types with quite a different
physical sense. The first type of diagrams -- like those in Fig.9 --
corresponds to the case, where all interactions with vacuum proceeds out
of the loop. Such diagrams correspond to logarithm corrections
(evolution) to the corresponding non-loop diagrams (without hard gluon
line). Since, as was discussed in Sec.3, we will not take into account
these  non-loop diagrams, then it seems reasonable, that at the same level
of accuracy we do not take into account their evolution. So, all the
diagrams of this type should be omitted. The problem of correct
calculation of non-loop diagrams and their leading logarithmic
correction is a special problem, which will not be discussed here. In
any case, estimations and physical reason show, that their contribution
would be significant at large $x$ and negligible in intermediate
region. We shall see at the end of the paper, that sum rules
themeselves indicate region of $x$ where effects of the non-loop
diagrams and their evolution may be neglected.

So, according to the previous discussion, we should bear in mind only
those diagrams, where interaction with vacuum takes place inside the
loop. (Figs.10,11). Such diagrams cannot be treated as evolution of any
non-loop diagrams and are pure power correction of dimension 6. All
these diagrams are, obviously, proportional to

$$
\langle 0\mid\bar{\psi}^d_{\alpha} ~ \psi^b_{\beta} D_{\rho} G^n_{\mu
\nu}\mid 0\rangle; ~~~ \langle 0\mid\bar{\psi}_{\alpha}^d
(\nabla_{\tau}\psi_{\beta}^b) G^n_{\mu \nu}\mid 0\rangle;
$$

$$
\langle 0 \mid (\nabla_{\tau} \bar{\psi}^d_{\alpha}) \psi_{\beta}^b G^n_{\mu
\nu} \mid 0\rangle
$$
These tensor structures were considered in \cite{9}
where using the equation of motion the following results have been obtained

%37
$$\langle 0 \mid
\bar{\psi}^d_{\alpha}\psi^b_{\beta}(D_{\sigma}G_{\mu\nu})^n \mid 0 \rangle =
\frac{g\langle 0 \mid \bar{\psi}\psi \mid 0 \rangle ^2}{3^3\cdot 2^5} \Biggl
(g_{\sigma\nu}\gamma^{\mu} - g^{\sigma\mu}\gamma_{\nu}\Biggr
)_{\beta\alpha}(\lambda^n)^{bd}$$
\be
\langle 0 \mid \bar{\psi}^d_{\alpha}
(\nabla_{\sigma}\psi_{\beta})^b G^n_{\mu\nu}\mid 0 \rangle =
\frac{g\langle 0 \mid \bar{\psi}\psi \mid 0 \rangle^2}{3^3\cdot 2^6}
[g^{\sigma\mu}\gamma_{\nu} - g^{\sigma\nu}\gamma_{\mu} -
i\varepsilon^{\sigma\mu\nu\lambda}\gamma^5\gamma_{\lambda}]_
{\beta\alpha}(\lambda^n)^{bd}
\label{37}
\ee
$\langle 0 \mid
(\nabla_{\sigma}\bar{\psi}_{\alpha})^d \bar{\psi}_{\beta}^bG_{\mu\nu}^n \mid
0 \rangle$ can be easily calculated using the results of \cite{9}

$$\langle 0 \mid
(\nabla_{\sigma}\bar{\psi}_{\alpha})^d\bar{\psi}_{\beta} ^bG_{\mu\nu}^n \mid
0 \rangle =
\frac{g\langle 0 \mid \bar{\psi}\psi \mid 0 \rangle^2}{3^3\cdot 2^6}
[g^{\sigma\mu}\gamma_{\nu} - g^{\sigma\nu}\gamma_{\mu} +
i\varepsilon^{\sigma\mu\nu\lambda}\gamma^5\gamma_{\lambda}]_
{\beta\alpha}(\lambda^n)^{bd}$$
For diagrams in Fig.10  we use the following expansion of gluon propagator

%38
$$
S^{np}_{\nu\rho}(x-y,y) = \frac{-i}{(2\pi)^4} gf^{npl}\int \frac{d^4
k}{k^4}e^{-iku} \cdot \left \{ \Biggl [
-ik_{\lambda}y_{\alpha}G_{\alpha\lambda}^l - \frac{2}{3}i
 (y_{\alpha}y_{\beta}k_{\lambda} -\right.$$
\be
\left.- \frac{i y_{\beta}}{k^2} (k^2\delta_{\alpha \lambda} -
2k_{\alpha}k_{\lambda})) (D_{\alpha} G_{\beta \lambda})^l +
\frac{1}{3}\Biggl (y_{\alpha} + \frac{2ik_{\alpha}}{k^2}\Biggr )(D_{\lambda}
G_{\alpha \lambda})^l\Biggr ] \delta_{\nu\rho} + 2 \Biggl [ G^l_{\nu\rho} +
2i \frac{k_{\alpha}}{k^2}(D_{\alpha}G_{\nu\rho})^b\Biggl ] \right \}
\label{38}
\ee
This expression can be found  using the a method of calculation of gluon
propagator in external vacuum gluon field, suggested in \cite{12}. The
same result up to  $\sim G$   term is explicitly written in \cite{13} (see
also \cite{17}).
 The total number of $d=6$ diagrams is enormous -- about 500.
 All of them were calculated by use of
REDUCE program. The final result for  d=6 correction after double Borel
transformation have the form

%39
$$Im~ \Pi^{d=6} = -\frac{1}{(2\pi)^7}\cdot g^2 \cdot
\frac{(ga)^2}{M^4}\cdot \frac{1}{3^6\cdot 2^5}\cdot \Biggl [ (-5784 x^4 -
1140 x^3
-20196 x^2 + 20628 x - 8292) ~ln(2) + $$
\be
+ 4740 x^4
+ 8847 x^3 + 2066 x^2 - 2553 x + 1416 \Biggr ] \frac{1}{x(1-x)^2}
\label{39}
\ee
$(ga)^2\equiv 4\pi\alpha_s \cdot (2\pi)^4\langle 0\mid
\bar{\psi}\psi\mid 0\rangle^2$.
Before we write  the final result of the sum rules, let us
make one note.  One can see, that in contribution of $d=6$ operator
(\ref{39}) strong coupling constant $g^2$ appears as factor, and again
it appears in structures  $(ga)^2$.  The factor $g^2$   corresponds to
interaction with quark propagator (vertices of hard gluon line in diagrams
in Fig.9,10, or vertices of external gluon in diagrams in Fig.6,7), and it
is reasonable to take it at the renormalization point  $\mu^2=Q^2_0$.
On the other side, $g^2$   in structure $(ga)^2$ appears as a consequence of
use of equation of motion and its normalization point should be taken
in such a way that the quantity $\alpha_s\langle 0\mid \bar{\psi}\psi \mid 0
\rangle^2$  is renormalization group invariant.
Finally, substituting results for bare loop (\ref{30}) and power corrections
(\ref{39}), we can write the sum rule for quark distribution
function in pion:

%40
$$x u_{\pi}(x) = \frac{3}{2\pi^2} \frac{M^2}{f^2_{\pi}}x^2(1-x)\cdot \Biggl
[\Biggl ( 1 + \Bigg ( \frac{\alpha_s(M^2)\cdot ln(Q^2_0/M^2)}{3\pi}
\Biggr ) \Biggl (\frac{1+4x~ln(1-x)}{x} - \frac{2(1-2x)ln~x}{1-x}\Biggr )
\Bigg )
$$
\be
\cdot (1-e^{-s_0/M^2}) - \frac{4\pi\alpha_s(Q^2_0)\cdot 4\pi
\alpha_s(M^2)a^2(M^2)}{(2\pi)^4\cdot 3^7\cdot 2^6 \cdot M^6}\cdot
\frac{\omega(x)}{x^3(1-x)^3}\Biggr ]
\label{40}
\ee
where $\omega(x)$ is the expression in square brackets in (\ref{39}).
We choose the effective LO QCD parameter $\Lambda_{QCD}=200~ MeV$,
$Q^2_0=2~GeV^2$.  The value of renorminvariant parameter
is equal

$$\alpha_sa^2 = \alpha_s(M^2 = 1~GeV^2)\cdot (0.55~GeV^3)^2=0.13~GeV^6$$
The value of $a$ was taken from the
best fit \cite{18} of the sum rule of nucleon masses (see \cite{9}, Appendix
B).  Continuum threshold was varied in the interval, $0.8 <s_0<1.2~GeV^2$
and it was found that the results only slightly depend on it.  The analysis
of the sum rule (\ref{40}) shows, that  they are fulfilled in the region
$0.15<x<0.7$; power corrections are less than 30\%, and continuum
contribution is small ($< 25\%$).  Stability in Borel mass parameter $M^2$
dependence in the region $0.4~ GeV^2 < M^2   < 0.6~ GeV^2$ is good;
especially in the region of $x\leq 0.4$ the $M^2$ dependence is almost
constant (see Fig. 12).

The final result for $u_{\pi}(x)$, (at $M^2=0.45~GeV^2$, $S_0=0.8~GeV^2$), is
shown in Fig.13 (thick solid line).
On Fig.13 is also plotted the curve of $u$-quark distribution in pion, found
in \cite{19} by using evolution equation  and the fit to the data on
Drell-Yan process, performed in  \cite{20}).  Bearing in mind, that
NLO $\alpha_s$-corrections are not accounted and one may expect, that they
would increase $u_{\pi}(x)$ at low $x$ and decrease at large $x$, one may
consider the agreement as good.  We also show in the same figure pure bare
loop contribution  (line with squares) and contribution (\ref{30}) of bare
loop with perturbative correction (crossed line).  One can see, that pure
bare loop is not in a quite good agreement with experiment and both
perturbative correction and power correction improve the agreement with
experiment.  Let us  discuss, why stability became worse when x became
larger (see Fig.12).  From our point of view, it reflect the influence  of
non-loop diagrams (and their evolution), which were not accounted as it was
discussed in sect.4.  Indeed, the non-loop diagram which formally are
proportional to $\delta(1-x)$, of course really would correspond to some
function with maximum close to x=1 and fast decreasing when $x$ decreased.
 That is why effects of such diagrams (and their evolution) are  negligible
 at $x\la 0.4-0.5$, but may be more or less sensible at large $x$, and
deterioration of stability probably reflects this fact. We repeat, finally
that obtained valence $u$-quark distribution function $u_{\pi}(x)$ can be
used as input for evolution equation (starting from point $Q^2_0=2~GeV^2$).

Let us now discuss at the end the estimations for the moments of quark
distribution which can be found with the help of the results obtained.

To get the moments, one should make some suggestions about the region of
small $x \la 0.15$ and large $x \ga 0.7 \div 0.8$, where our method is
inapplicable. Of course, in this case the estimation of moments are not
model-independent and the accuracy of estimation of moments should be
treated as lower than for the structure function (\ref{40}) itself.

 If we make a natural supposition, that at $x \la 0.15~u_{\pi}(x) \sim
 1/\sqrt{x}$  according to Regge behaviour, and at large $x \ga 0.7,~
 u_{\pi} (x) \sim (1 -x)^2$ according to quark counting rules, then,
 matching these functions with our result (\ref{40}), one may find, that

$$
{\cal{M}}_0 = \int\limits^{1}_{0} u_{\pi}(x) dx \approx 0.84
$$

$$
{\cal{M}}_1 =  \int\limits^{1}_{0} xu_{\pi}(x) dx \approx 0.21
$$
at $M^2 = 0.4 \div 0.45 GeV^2$.
These results only slightly depend on the choice of the points of matching
(not more than 5\% when we vary lower matching point in the region $0.15
\div 0.2$ and the upper one in the region $0.65 \div 0.75$. One may note
that ${\cal{M}}_0$ which has the physical meaning of the number of quarks
in pion (and should be ${\cal{M}}_0 = 1)$ is really close to 1 within our
accuracy $\sim 10 \div 20\%$.  ${\cal{M}}_1$ has physical meaning of the part
of momentum carried by a valence quark, and the value ${\cal{M}}_1 \approx
0.21$ is in good agreement with well known fact, that \underline{two}
valence quarks in pion take about 40\% of the total momentum.

This work was supported in part by RFBR grant 97-02-16131.

\newpage

\centerline{\bf Appendix}

\setcounter{equation}{0}

\def\theequation{A.\arabic{equation}}

\bigskip

In this Appendix the double dispersion representation (\ref{22}) of integral
(\ref{19})  is proved. It is convenient to change variables in (\ref{19})
and put  $p-k=k^{\prime}$. Then (\ref{19}) takes the form (prime is omitted)

%a1
\be
Im~T(p^2_1,p^2_2,q^2,\nu) = \int d^4 k \frac{1}{(p_1-k)^2}
\frac{1}{(p_2-k)^2} \delta[(p+q-k)^2] \delta(k^2)
\label{A.1}
\ee
Assume  that $q^2=q^{\prime 2}$,  $t=(p_1-p_2)^2=0$  and choose the Lorenz
system, where 4-vector $P=(p_1+p_2)/2$  has only $z$-component equal $P_z$.
From

%A2
\be
t=(p_1-p_2)^2 = p^2_1 + p^2_2 + 2p_1 p_2 =0
\label{A.2}
\ee
it follows

%A3
\be
P^2 = -P^2_z = (p^2_1+p^2_2)/2
\label{A.3}
\ee
Introduce 4-null-vector $r=p_1-p_2$, $r^2=0$

%a4
\be
rP=(p^2_1 - p^2_2)/2
\label{A.4}
\ee
We have

%a5
\be
p_1 = P + r/2, ~~~~~p_2 = P -r/2
\label{A.5}
\ee
Use the notation

%a6
\be
qP = \nu = q p_1 = q p_2
\label{A.6}
\ee
Then

%a7
\be
q^{\prime}p_1 = q^{\prime}p_2 = \nu + (p^2_1 - p^2_2)/2, ~~~qr = q^{\prime}r
=0
\label{A.7}
\ee
We can choose the coordinate system, where 4-vector $q$ has only time and
$z$-components and

%a8
\be
q_z = -\nu/\sqrt{-P^2},~~~~~~~q_0 = \sqrt{\nu^2-q^2P^2}/\sqrt{-P^2} \approx
\nu/\sqrt{-P^2} =-q_z
\label{A.8}
\ee
The last equality corresponds to account of lower twist terms. From
(\ref{A.4}) and (\ref{A.7})  for 4-vector $r$ with components
$r=\{r_0,r_{\perp},r_z\}$  it follows

%a9
$$
r_0 = \frac{1}{2}\nu
\frac{p^2_1-p^2_2}{\sqrt{-P^2}}\frac{1}{\sqrt{\nu^2-q^2P^2}}\approx
\frac{1}{2}~~\frac{p_1 - p^2_2}{\sqrt{-P^2}},~~ r_z = \frac{1}{2}~~
\frac{p^2_1 - p^2_2}{\sqrt{-P^2}} \approx r_0,
$$
\be
r^2_{\bot}= -\frac{1}{4}~
q^2 \frac{(p^2_1 - p^2_2)^2}{\nu^2 - q^2P^2}
\label{A.9}
\ee
The components $r_0$ and $r_z$ are equal in the lowest twist approximation
and of order $\sqrt{-P^2}$ if $p^2_1 \sim p^2_2$, while $r_{\perp}\sim
(p^2_1 - p^2_2)/\nu^{1/2}$, i.e. of the next order in this approximation and
may be neglected. The argument of the first $\delta$-function in (\ref{A.1})
is equal

%a10
\be
s - 2~ \frac{1}{\sqrt{-P^2}}\Biggl [\nu + P^2 + \frac{1}{4}(p^2_1 - p^2_2)
\Biggr ]k_z - \Biggl [2 \sqrt{\frac{\nu^2 - q^2P^2}{-P^2}} + \frac{1}{2}
~~ \frac{p^2_1 - p^2_2}{\sqrt{-P^2}} \Biggr ]k_0 + r_{\bot} k_{\bot} Cos
\varphi = 0
\label{A.10}
\ee
where $\varphi$ is the azimutal angle between $p_{\perp}$ and $k_{\perp}$.
The the last   term in (\ref{A.10})  may be
omitted -- it is of the next order in $p^2/\nu$: it may appear only
squared because of integration over $\varphi$ in (\ref{A.1})). (This
fact can be proved by direct calculation.) From the inequality

%a11
\be
k^2_0 - k^2_z \geq 0,
\label{A.11}
\ee
the inequality follows, which defines the integration  domain over $k_z$ in
the integral (\ref{A.1}):

%a12
\be
k^2_zP^2 - \sqrt{-P^2}[\nu +P^2 + \frac{1}{4}(p^2_1 - p^2_2)] k_z -
\frac{1}{4}P^2 \geq
\label{A.12}
\ee
It is convenient to use the notation

%a13
\be
\upsilon = 2\sqrt{-P^2}k_z
\label{A.13}
\ee
The integration domain is

%a14
\be
- 2\nu  \leq \upsilon  \leq -P^2(1-x)
\label{A.14}
\ee
The denominators in (\ref{A.1}) are calculated by using the relations:

%a15
\be
(p_1-k)^2 = p^2_1 - 2p_1k = p^2_1 - 2Pk - rk ~~~(p_2-k)^2 = p^2_2 - 2Pk + rk
\label{A.15}
\ee

%a16
\be
Pk = -\sqrt{-P^2}k_z,~~~~rk \approx r_0(k_0+k_z) =
r_0\sqrt{-P^2}(1-x)=\frac{1}{2}(p^2_1 - p^2_2)(1-x)
\label{A.16}
\ee
(In the above equalities (\ref{A.10}) and (\ref{A.9}) were exploited). As a
result we get ($\delta$-functions were eliminated by integration over
$k^2_{\perp}$ and $k_0$):

%a17
\be
Im~T = \frac{\pi}{4q_0\sqrt{-P^2}}\int\limits^{-P^2(1-x)}_{-2\nu} d\upsilon
\frac{1}{p^2_1+\upsilon -(p^2_1-p^2_2)(1-x)/2}\times
\frac{1}{p^2_2+\upsilon +(p^2_1-p^2_2)(1-x)/2}
\label{A.17}
\ee
Changing variables

%a18
\be
\upsilon=-P^2(1-x) - ux
\label{A.18}
\ee
gives the final answer

%a19
\be
Im~T = \frac{\pi}{4\nu x}\int\limits^{\infty}_0
\frac{1}{u-p^2_1}\frac{1}{u-p^2_2}du
\label{A.19}
\ee
(The upper limit of integration was put as infinity, what is legitimate in
the lowest twist approximation).

\newpage

\begin{figure}
\epsfxsize=10cm
\epsfbox{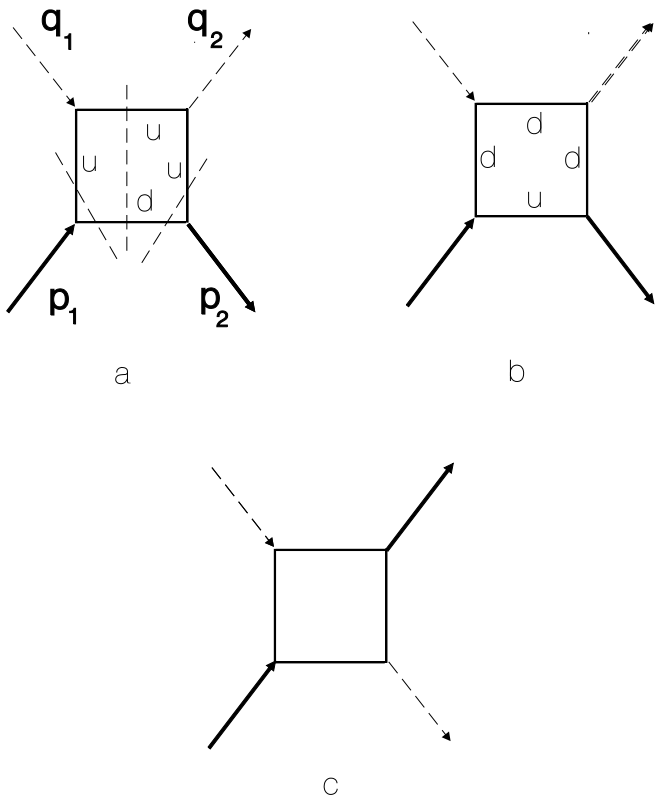}
\caption{Diagrams, corresponding to unit operator contribution. Dashed 
lines with arrows correspond to photon, thick solid - to hadron current}
\end{figure}
\newpage

\begin{figure}
\epsfxsize=10cm
\epsfbox{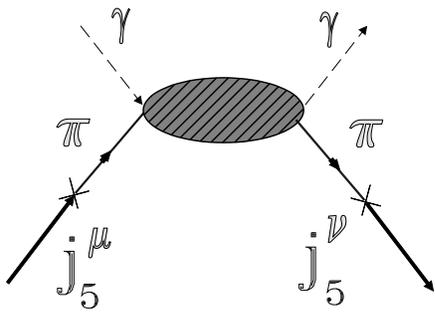}
\caption{Diagram of forward photon-pion scattering.}
\end{figure}
\newpage

\begin{figure}
\epsfxsize=10cm
\epsfbox{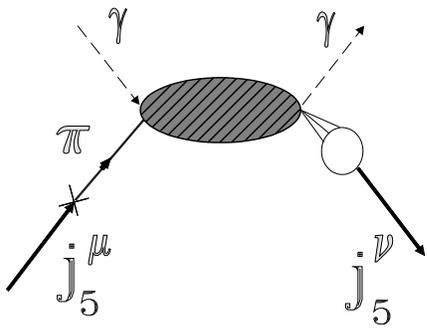}
\caption{Example of non-diagonal transition}
\end{figure}
\newpage

\begin{figure}
\epsfxsize=10cm
\epsfbox{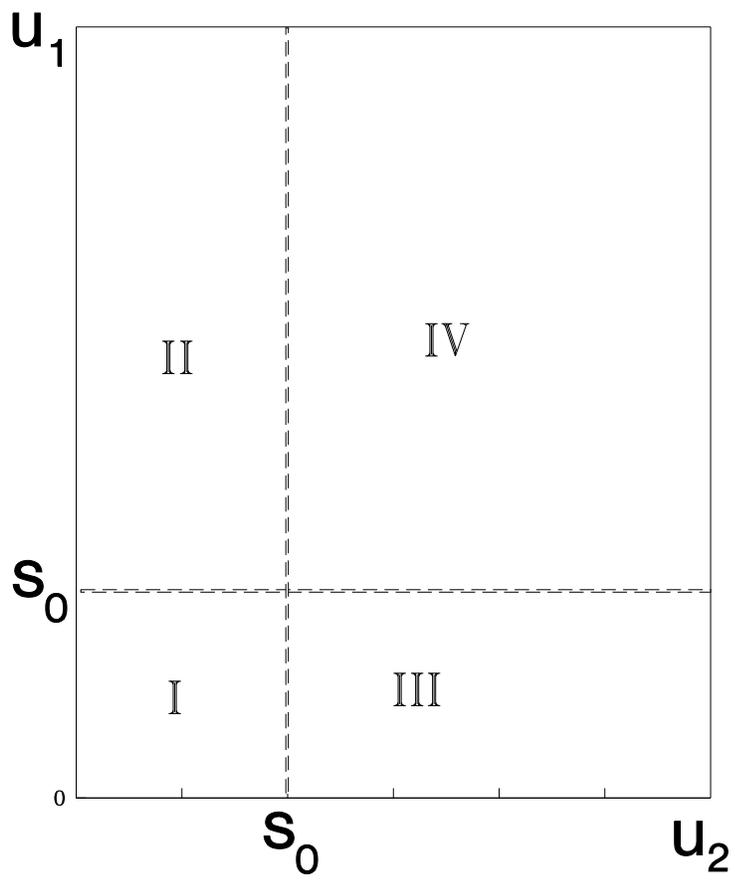}
\caption{Integration region in double dispersion representation.}
\end{figure}
\newpage

\begin{figure}
\epsfxsize=10cm
\epsfbox{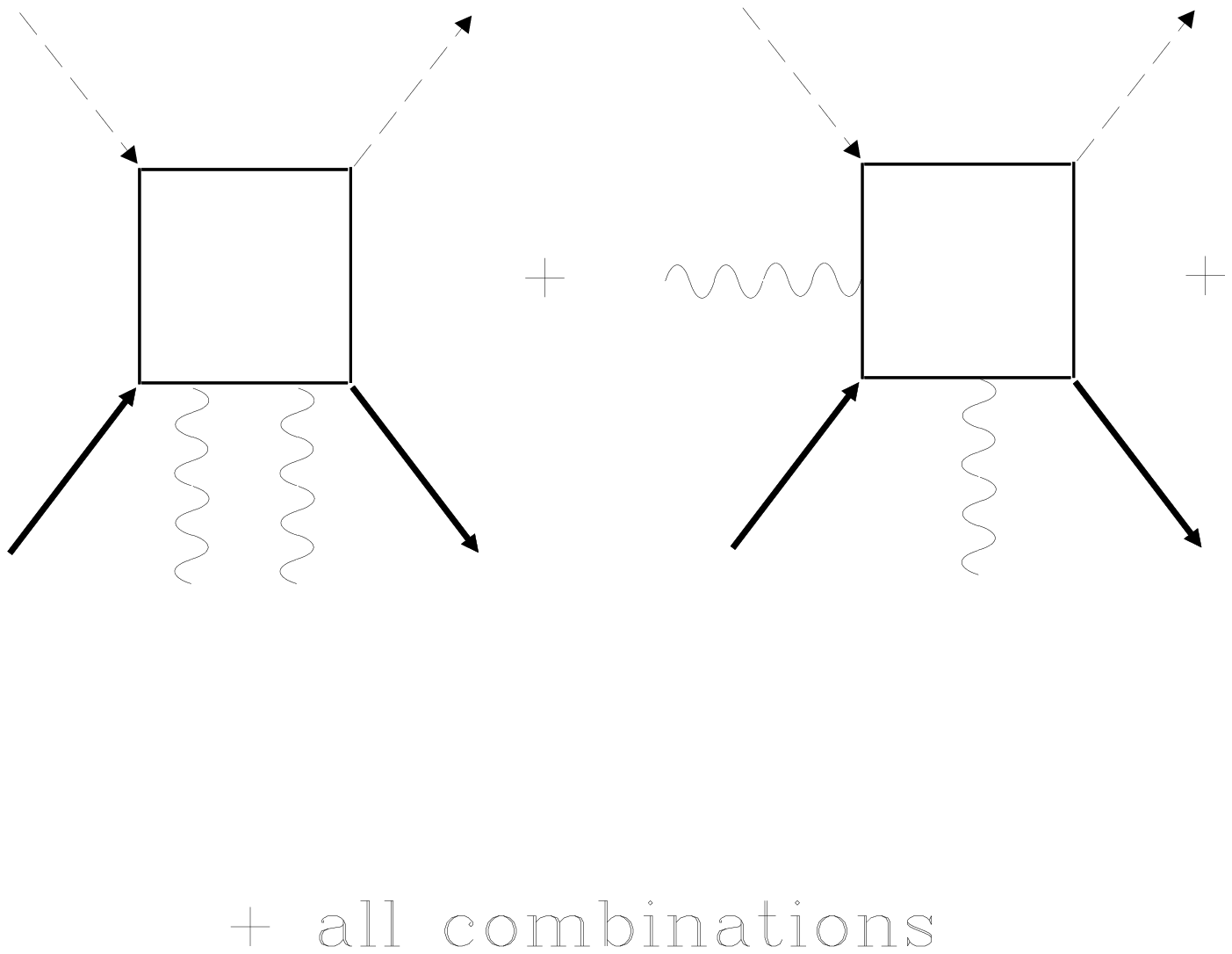}
\caption{Diagrams, corresponding to $d=4$ operator contribution. Dashed 
lines with aroows correspond to photon, thick solid - hadron current,
wave lines correspond to external gluon field}
\end{figure}
\newpage

\begin{figure}
\epsfxsize=10cm
\epsfbox{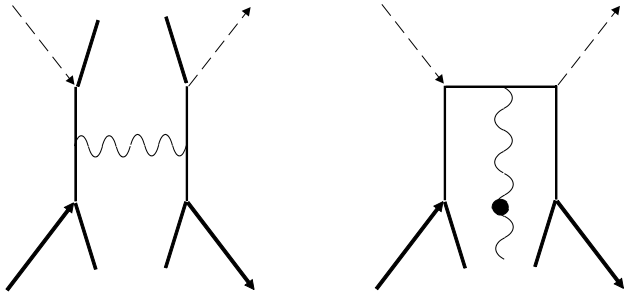}
\caption{Examples of non-loop diagrams of dimension 4. Wave lines corespond 
to gluons, dot means derivative, other notations as in Fig.1}
\end{figure}
\newpage

\begin{figure}
\epsfxsize=10cm
\epsfbox{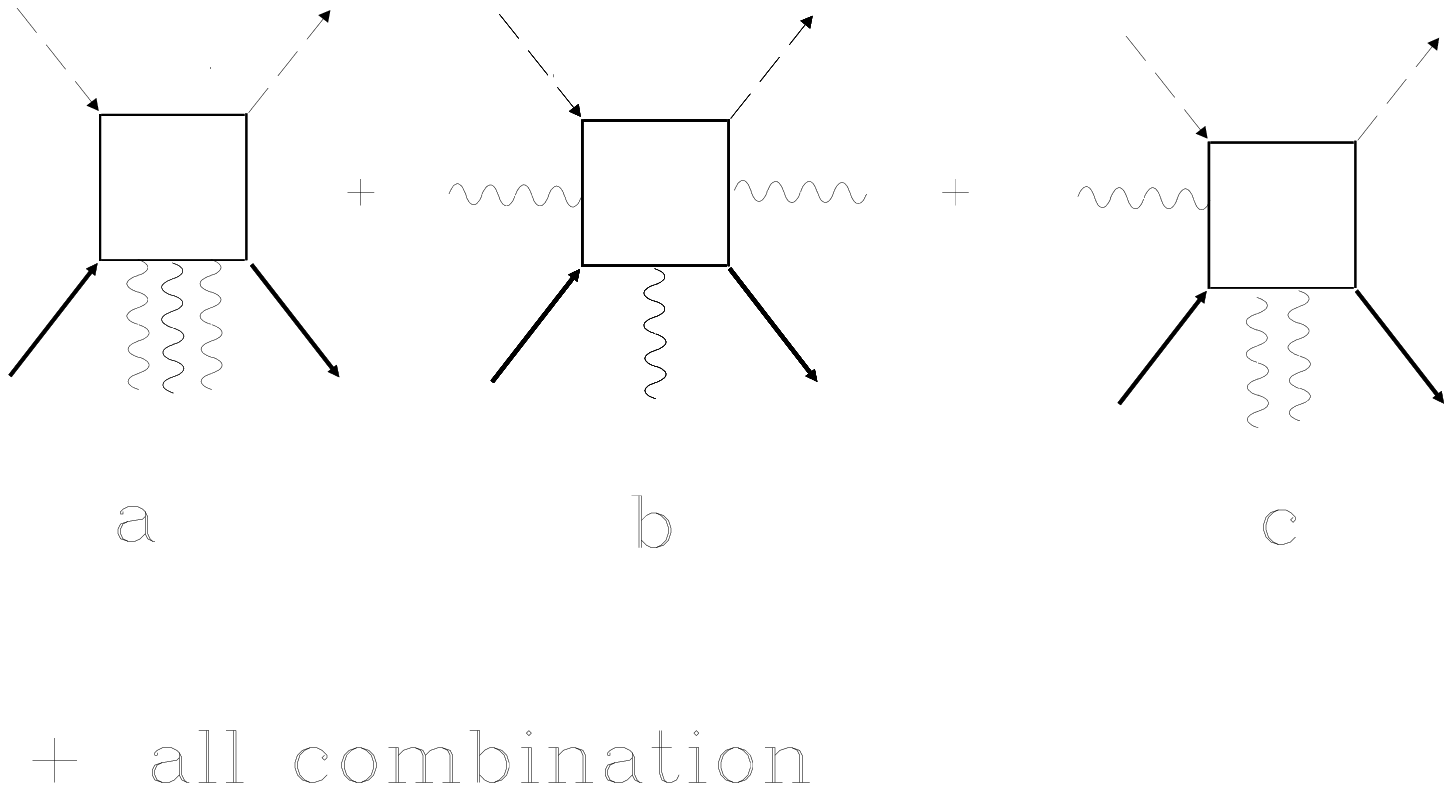}
\caption{Diagrams of dimension 6, see text. All notations as in Fig.6}
\end{figure}
\newpage

\begin{figure}
\epsfxsize=10cm
\epsfbox{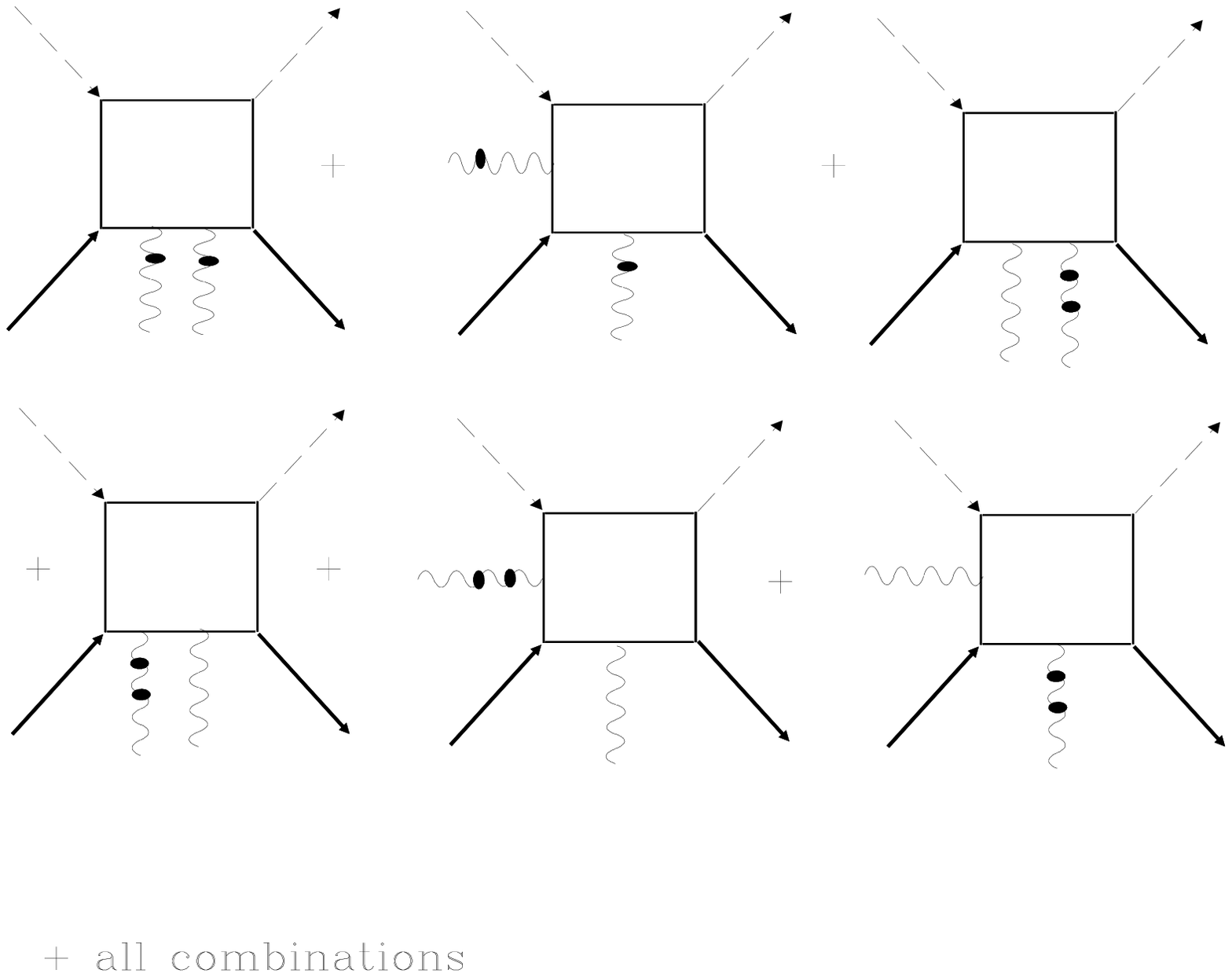}
\caption{Diagrams of dimension 6. External gluon line with 
dot corresponds to derivatives of gluon lines. All notations as in Fig.6}
\end{figure}
\newpage

\begin{figure}
\epsfxsize=10cm
\epsfbox{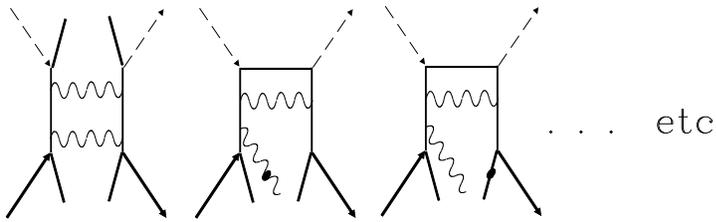}
\caption{Dimension 6 giagrams with out the loop vacuum interaction. 
All notations as in Fig.8}
\end{figure}
\newpage

\begin{figure}
\epsfxsize=10cm
\epsfbox{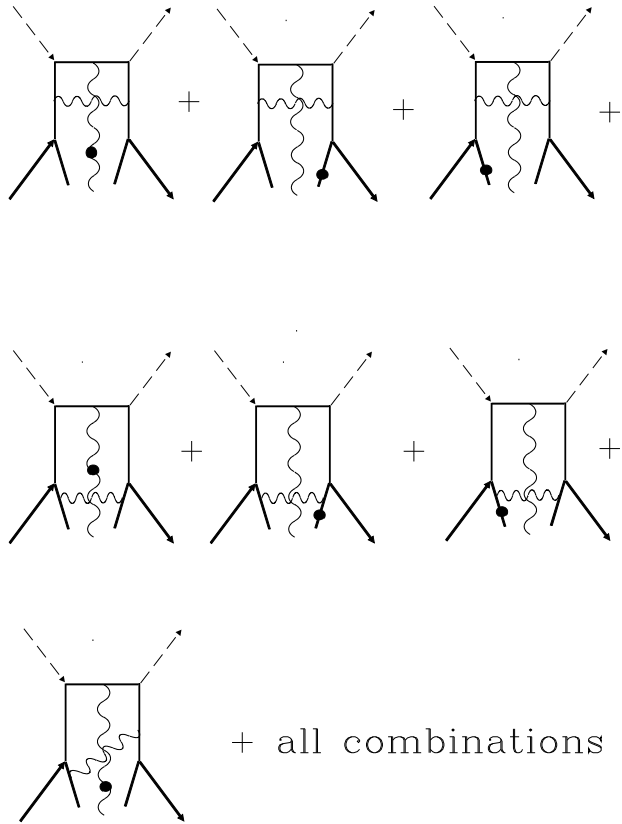}
\caption{Diagrams of dimension 6, corresponding to quark propogator 
expansion (eq.37). All notations as in Fig.8}
\end{figure}
\newpage

\begin{figure}
\epsfxsize=10cm
\epsfbox{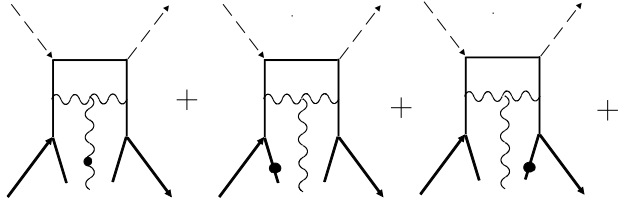}
\caption{Diagrams of dimension 6, corresponding to quark and gluon 
propogator expansion (eq.37, 38). All notations as in Fig.8}
\end{figure}
\newpage

\begin{figure}
\epsfxsize=10cm
\epsfbox{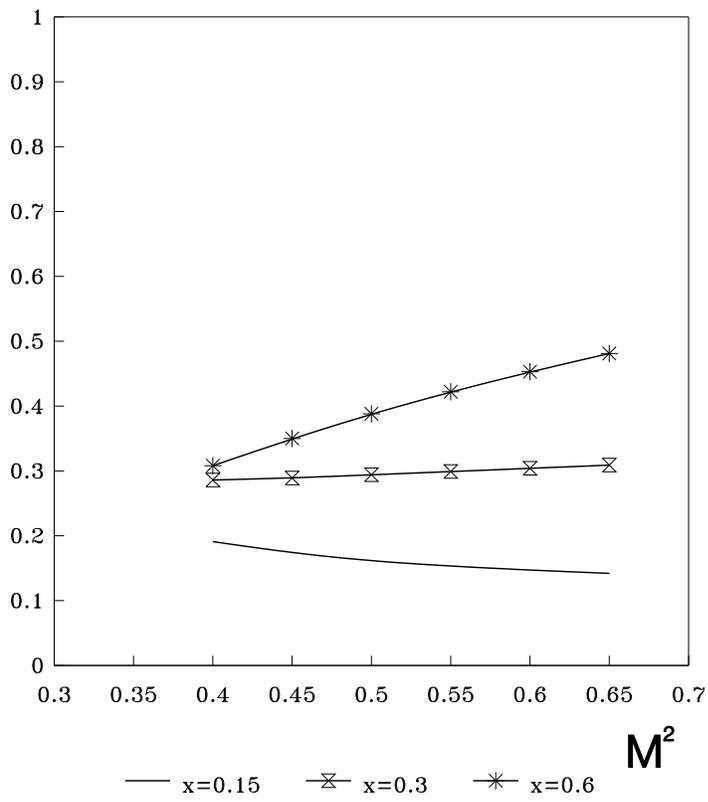}
\caption{Borel mass dependence of quark distribution function at various $x$}
\end{figure}
\newpage

\begin{figure}
\epsfxsize=10cm
\epsfbox{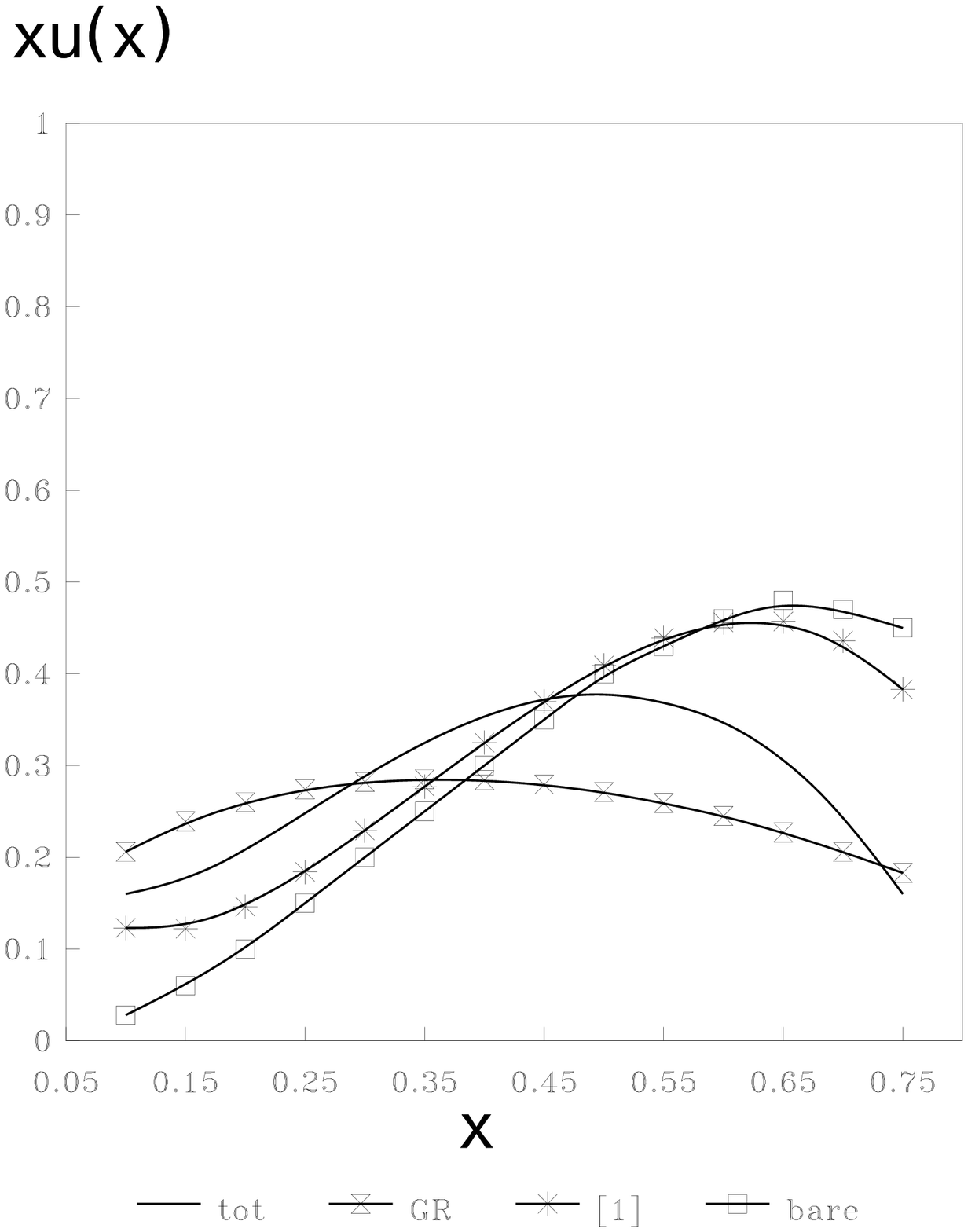}
\caption{Quark distribution function in pion, noted "total". For comparison
fit from [19], noted "GR", is shown. Also bare loop ("bare") and bare loop 
with perturbative corrections (noted "1"), are shown}
\end{figure}
\newpage


\newpage

\begin{thebibliography}{99}
\bibitem{1} M.A.Shifman, A.I.Vainshtein and V.I.Zakharov, Nucl. Phys.
{\bf B147} (1979) 385, 448.
\bibitem{2}  Vacuum Structure and QCD Sum Rules,
ed. by M.Shifman, North Holland, Amsterdam, 1992.
\bibitem{3} Ya.Ya.
Balitsky Sov.J.Nucl.Phys., {\bf 37} (1983) 576.
\bibitem{4} Kolesnichenko
A.V. Yad.Fiz., {\bf 39} (1982) 1532.
\bibitem{5} Belyaev  V.M. and  Block B.
Yu., Yad.Fiz., {\bf 43} (1986) 706; Phys.Lett. 167B (1986) 99.
\bibitem{6}  B.L.Ioffe, Pisma ZhETF,
{\bf 42} (1985) 266, {\bf 43} (1986) 316.
\bibitem{7} Belyaev V.M. and Ioffe
B.L., Nucl. Phys., {\bf B310} (1988) 548.
\bibitem{8}  A.S.Gorsky,
B.L.Ioffe, A.Yu.  Khodjamirian and  A.Oganesian Z.Phys.C, {\bf 44} (1989)
523.
\bibitem{9} B.L.Ioffe and A.V.Smilga, Nucl.Phys. B232 (1984) 109.
\bibitem{10} B.L.Ioffe, Phys.At.Nucl., {\bf 58} (1995) 1492.
\bibitem{11} B.L.Ioffe and A.V.Smilga, Nucl.Phys. B216 (1983) 373.
\bibitem{12} A.V.Smilga, Yad.Fiz. {\bf 35} (1982) 473.
\bibitem{13}  V.A.Novikov,
M.A.Shifman, A.I.Vainshtein and  V.I.Zakharov, Fortschr.Phys. {\bf 32} (1984)
585.
\bibitem{14} S.N.Nikolaev and  A.V.Radyushkin, Nucl.Phys. {\bf B213} (1983)
189.
\bibitem{15} V.A.Novikov, M.A.Shifman, A.I.Vainstein and V.I.Zakharov,
Phys.Lett. {\bf 86B} (1979) 347.
\bibitem{16} T.Sch\"afer and E.V.Shuryak, Rev.Mod.Phys.,
\bibitem{17} J.Govaerts, F.de Viron, D.Gusbin and J.Weyers, Nucl.Phys., {\bf
B248} (1984) 1.
\bibitem{18} B.L.Ioffe and A.G.Oganesian, Phys.Rev., {\bf D57} (1998) R6590.
\bibitem{19} M.Gluck , E.Reya and  A Vogt Z.Phys.C,  {\bf 53},
651-655 (1992).
\bibitem{20} P.Aurenche et al, Phys.Lett., {\bf B233} (1989), 517.


\end{thebibliography}
\end{document}